\documentclass[aps,prl,superscriptaddress,showpacs,floatfix,twocolumn]{revtex4}

\usepackage{graphicx}   

\begin{document}

\title{Measurement of identified $\pi^0$ and inclusive photon $v_2$
and implication to the direct photon production in
$\sqrt{s_{_{NN}}}$ = 200 GeV Au+Au collisions}

\newcommand{\abilene}{Abilene Christian University, Abilene, TX 79699, USA}
\newcommand{\acadsin}{Institute of Physics, Academia Sinica, Taipei 11529, Taiwan}
\newcommand{\banaras}{Department of Physics, Banaras Hindu University, Varanasi 221005, India}
\newcommand{\barc}{Bhabha Atomic Research Centre, Bombay 400 085, India}
\newcommand{\bnl}{Brookhaven National Laboratory, Upton, NY 11973-5000, USA}
\newcommand{\caucr}{University of California - Riverside, Riverside, CA 92521, USA}
\newcommand{\ciae}{China Institute of Atomic Energy (CIAE), Beijing, People's Republic of China}
\newcommand{\cns}{Center for Nuclear Study, Graduate School of Science, University of Tokyo, 7-3-1 Hongo, Bunkyo, Tokyo 113-0033, Japan}
\newcommand{\columbia}{Columbia University, New York, NY 10027 and Nevis Laboratories, Irvington, NY 10533, USA}
\newcommand{\dapnia}{Dapnia, CEA Saclay, F-91191, Gif-sur-Yvette, France}
\newcommand{\debrecen}{Debrecen University, H-4010 Debrecen, Egyetem t{\'e}r 1, Hungary}
\newcommand{\fsu}{Florida State University, Tallahassee, FL 32306, USA}
\newcommand{\gsu}{Georgia State University, Atlanta, GA 30303, USA}
\newcommand{\hiroshima}{Hiroshima University, Kagamiyama, Higashi-Hiroshima 739-8526, Japan}
\newcommand{\ihepprot}{IHEP Protvino, State Research Center of Russian Federation, Institute for High Energy Physics, Protvino, 142281, Russia}
\newcommand{\isu}{Iowa State University, Ames, IA 50011, USA}
\newcommand{\jinrdubna}{Joint Institute for Nuclear Research, 141980 Dubna, Moscow Region, Russia}
\newcommand{\kaeri}{KAERI, Cyclotron Application Laboratory, Seoul, South Korea}
\newcommand{\kangnung}{Kangnung National University, Kangnung 210-702, South Korea}
\newcommand{\kek}{KEK, High Energy Accelerator Research Organization, Tsukuba, Ibaraki 305-0801, Japan}
\newcommand{\kfki}{KFKI Research Institute for Particle and Nuclear Physics of the Hungarian Academy of Sciences (MTA KFKI RMKI), H-1525 Budapest 114, POBox 49, Budapest, Hungary}
\newcommand{\korea}{Korea University, Seoul, 136-701, Korea}
\newcommand{\kurchatov}{Russian Research Center ``Kurchatov Institute", Moscow, Russia}
\newcommand{\kyoto}{Kyoto University, Kyoto 606-8502, Japan}
\newcommand{\labllr}{Laboratoire Leprince-Ringuet, Ecole Polytechnique, CNRS-IN2P3, Route de Saclay, F-91128, Palaiseau, France}
\newcommand{\lawllnl}{Lawrence Livermore National Laboratory, Livermore, CA 94550, USA}
\newcommand{\losalamos}{Los Alamos National Laboratory, Los Alamos, NM 87545, USA}
\newcommand{\lpc}{LPC, Universit{\'e} Blaise Pascal, CNRS-IN2P3, Clermont-Fd, 63177 Aubiere Cedex, France}
\newcommand{\lund}{Department of Physics, Lund University, Box 118, SE-221 00 Lund, Sweden}
\newcommand{\muenster}{Institut f\"ur Kernphysik, University of Muenster, D-48149 Muenster, Germany}
\newcommand{\myongji}{Myongji University, Yongin, Kyonggido 449-728, Korea}
\newcommand{\nagasaki}{Nagasaki Institute of Applied Science, Nagasaki-shi, Nagasaki 851-0193, Japan}
\newcommand{\newmex}{University of New Mexico, Albuquerque, NM 87131, USA}
\newcommand{\nmsu}{New Mexico State University, Las Cruces, NM 88003, USA}
\newcommand{\ornl}{Oak Ridge National Laboratory, Oak Ridge, TN 37831, USA}
\newcommand{\orsay}{IPN-Orsay, Universite Paris Sud, CNRS-IN2P3, BP1, F-91406, Orsay, France}
\newcommand{\pnpi}{PNPI, Petersburg Nuclear Physics Institute, Gatchina,  Leningrad region, 188300, Russia}
\newcommand{\riken}{RIKEN, The Institute of Physical and Chemical Research, Wako, Saitama 351-0198, Japan}
\newcommand{\rikjrbrc}{RIKEN BNL Research Center, Brookhaven National Laboratory, Upton, NY 11973-5000, USA}
\newcommand{\saispbstu}{Saint Petersburg State Polytechnic University, St. Petersburg, Russia}
\newcommand{\saopaulo}{Universidade de S{\~a}o Paulo, Instituto de F\'{\i}sica, Caixa Postal 66318, S{\~a}o Paulo CEP05315-970, Brazil}
\newcommand{\seoulnat}{System Electronics Laboratory, Seoul National University, Seoul, South Korea}
\newcommand{\stonybrkc}{Chemistry Department, Stony Brook University, SUNY, Stony Brook, NY 11794-3400, USA}
\newcommand{\stonycrkp}{Department of Physics and Astronomy, Stony Brook University, SUNY, Stony Brook, NY 11794, USA}
\newcommand{\subatech}{SUBATECH (Ecole des Mines de Nantes, CNRS-IN2P3, Universit{\'e} de Nantes) BP 20722 - 44307, Nantes, France}
\newcommand{\tenn}{University of Tennessee, Knoxville, TN 37996, USA}
\newcommand{\titech}{Department of Physics, Tokyo Institute of Technology, Tokyo, 152-8551, Japan}
\newcommand{\tsukuba}{Institute of Physics, University of Tsukuba, Tsukuba, Ibaraki 305, Japan}
\newcommand{\vandy}{Vanderbilt University, Nashville, TN 37235, USA}
\newcommand{\waseda}{Waseda University, Advanced Research Institute for Science and Engineering, 17 Kikui-cho, Shinjuku-ku, Tokyo 162-0044, Japan}
\newcommand{\weizmann}{Weizmann Institute, Rehovot 76100, Israel}
\newcommand{\yonsei}{Yonsei University, IPAP, Seoul 120-749, Korea}
\affiliation{\abilene}
\affiliation{\acadsin}
\affiliation{\banaras}
\affiliation{\barc}
\affiliation{\bnl}
\affiliation{\caucr}
\affiliation{\ciae}
\affiliation{\cns}
\affiliation{\columbia}
\affiliation{\dapnia}
\affiliation{\debrecen}
\affiliation{\fsu}
\affiliation{\gsu}
\affiliation{\hiroshima}
\affiliation{\ihepprot}
\affiliation{\isu}
\affiliation{\jinrdubna}
\affiliation{\kaeri}
\affiliation{\kangnung}
\affiliation{\kek}
\affiliation{\kfki}
\affiliation{\korea}
\affiliation{\kurchatov}
\affiliation{\kyoto}
\affiliation{\labllr}
\affiliation{\lawllnl}
\affiliation{\losalamos}
\affiliation{\lpc}
\affiliation{\lund}
\affiliation{\muenster}
\affiliation{\myongji}
\affiliation{\nagasaki}
\affiliation{\newmex}
\affiliation{\nmsu}
\affiliation{\ornl}
\affiliation{\orsay}
\affiliation{\pnpi}
\affiliation{\riken}
\affiliation{\rikjrbrc}
\affiliation{\saispbstu}
\affiliation{\saopaulo}
\affiliation{\seoulnat}
\affiliation{\stonybrkc}
\affiliation{\stonycrkp}
\affiliation{\subatech}
\affiliation{\tenn}
\affiliation{\titech}
\affiliation{\tsukuba}
\affiliation{\vandy}
\affiliation{\waseda}
\affiliation{\weizmann}
\affiliation{\yonsei}
\author{S.S.~Adler}	\affiliation{\bnl}
\author{S.~Afanasiev}	\affiliation{\jinrdubna}
\author{C.~Aidala}	\affiliation{\bnl}
\author{N.N.~Ajitanand}	\affiliation{\stonybrkc}
\author{Y.~Akiba}	\affiliation{\kek} \affiliation{\riken}
\author{J.~Alexander}	\affiliation{\stonybrkc}
\author{R.~Amirikas}	\affiliation{\fsu}
\author{L.~Aphecetche}	\affiliation{\subatech}
\author{S.H.~Aronson}	\affiliation{\bnl}
\author{R.~Averbeck}	\affiliation{\stonycrkp}
% ...
\collaboration{PHENIX Collaboration} \noaffiliation

\date{\today}

\begin{abstract}
The azimuthal distribution of identified $\pi^0$ and 
inclusive photons has been measured in $\sqrt{s_{NN}} = 200$~GeV 
Au+Au collisions with the PHENIX experiment at the Relativistic
Heavy Ion Collider (RHIC).  
The second harmonic parameter ($v_2$) was measured to 
describe the observed anisotropy of the azimuthal distribution. 
The measured inclusive photon $v_2$ is consistent with the value 
expected for the photons from
hadron decay and is also consistent with the lack of direct photon 
signal over the measured $p_T$ range 1-6 GeV/${\it c}$. 
An attempt is made to extract $v_2$ of direct photons.
\end{abstract}



\maketitle

Among the most exciting features of the experimental data from
the Relativistic Heavy-Ion Collider (RHIC) are 
the suppression of high $p_T$ hadron yields
\cite{phx200201,phx200301,phx200302,phx200401,str200201}, 
the baryon excess at intermediate $p_T$
\cite{phx200305,phx200402,phx200403,phx200404}, 
and the quark number scaling of the identified hadron 
$v_2$ \cite{phx200304,str200401}.   Theoretically, the observed 
high $p_T$ suppression has been attributed to energy loss of the 
hard-scattered partons~\cite{wang,GLV}.  
Experimentally, the absence of the suppression in d+Au 
collisions has shown that it is a final-state effect due 
to the hot and dense matter created in central Au+Au 
collisions \cite{phx200303,str200301,pbs200301,brm200301}. 
The quark number scaling of the measured elliptic 
flow parameter $v_2$ and the nuclear modification factor $R_{cp}$ 
of baryons versus mesons may suggest the existence of a thermalized
partonic phase before hadronization~\cite{Fries:Reco,Greco:Reco}. 

The second harmonic coefficient parameter $v_2$ of
the azimuthal distribution of the particles produced 
in heavy-ion collisions is defined by
\begin{equation}
\frac{dN}{d\phi} \propto 1 + 2 \; v_2 \cos(2(\phi-\Phi_{\rm RP})),
\end{equation}
\noindent
where $\phi$ is the azimuthal direction of the particle and 
$\Phi_{\rm RP}$ is the direction of the nuclear impact 
parameter (reaction plane) in a given collision. 
The $v_2$ in high-energy heavy-ion collisions is considered to 
be sensitive to the initial geometric overlap of the
colliding nuclei as well as the later expansion driven
by the initial pressure. 
Theoretically, the dominant source of $v_2$ at low $p_T$ is 
the expansion of the dense matter in the direction of the 
short axis of the overlap zone, and at high $p_T$ is 
the parton energy loss given by the shape of 
the geometrical overlap. The quark coalescence (recombination) 
might be responsible for the $v_2$ in the intermediate $p_T$ region. 
However, the experimental definition of $v_2$ includes any 
2nd harmonic correlation with respect to the event plane, which 
is given by the beam direction and the impact parameter 
direction.  
Detailed $v_2$ measurements of identified particles at
higher $p_T$ than 2~GeV/${\it c}$, where hydro-dynamics alone does not 
describe the measurements, would enable us to understand the different 
mechanisms that generate $v_2$ and to investigate the 
transition region from low to high $p_T$. Especially, the $v_2$ of 
identified $\pi^0$ will give a baseline measurement of 
inclusive photon $v_2$ to extract the direct photon $v_2$.

The direct photons produced in hard scattering are 
penetrating probes of the produced dense matter in 
heavy-ion collisions. 
Recently, we observed that the centrality dependence 
of the direct photon yield in $\sqrt{s_{NN}}$ = 
200~GeV Au+Au collisions is consistent 
with binary collision scaling \cite{phx200500}. 
The lack of suppression of direct photons is further evidence 
in favor of the final-state effect in hadron suppression.
In addition to the initially-produced hard photons that 
should inherently follow binary scaling, there may be
other counteracting effects resulting in apparent 
binary scaling.  For example, some fraction of the photons may 
originate from partons having experienced energy loss, 
causing an analogous suppression of these photons~\cite{suppr} 
similar to hadrons.  
On the other hand, 
the parton energy loss may enhance the photon yield 
via Bremsstrahlung while passing through 
the hot and dense matter~\cite{Fries:2003br}. 
The thermal emission of photons radiated from the hot 
and dense matter is also expected to increase direct 
photon yield for central Au+Au collisions~\cite{thermal}. 

The $v_2$ measurement of the direct photons
could help to confirm that the observed binary scaling 
of the direct photon excess is attributable to the direct
photon production being dominated by the initial hard scattering.
The $v_2$ measurement of the direct photons would give additional 
and complementary information to help disentangle the 
various scenarios of direct photon production, as well as 
to provide more information on the dynamics and properties 
of the produced hot and dense matter. The $v_2$ of 
photons from the initial Compton-like hard scattering 
is expected to be zero if they do not interact with 
the hot and dense matter produced during the collision. 
However when the $v_2$ of high $p_T$ hadrons is given purely 
by the parton energy loss, the photons from the 
parton fragmentation outside of the reaction zone should 
have $v_2$ similar to the hadrons at high $p_T$. Such 
photon fraction is expected to be about 50$\%$ of total 
direct photon yield at 3.5 GeV/${\it c}$ in 
$p_T$ \cite{suppr,Fries:2003br}. 
On the other hand, one would 
expect that the photons originating from Bremsstrahlung
due to the passage of partons through the hot and dense 
matter should have the opposite (negative) sign in $v_2$ 
compared with hadrons, because the parton energy loss 
is larger in the long axis of the overlapping region
(out-of-plane). Finally, the photons from the thermal 
radiation should reflect the dynamical evolution of 
the produced hot and dense matter. There are recent
theoretical predictions for different mechanisms \cite{turb}.

In this letter we present measurements of the $v_2$ of $\pi^0$ and
inclusive $\gamma$, as a function of transverse momentum and
collision centrality, and we discuss the implications for the
yield and $v_2$ of direct photons.  The data are for 200~GeV
Au+Au collisions from the PHENIX experiment \cite{phx200306}
recorded during Run-2 (2001) at RHIC.    The event trigger
and centrality definition are given by the Beam-Beam Counters 
(BBC) and the Zero Degree Calorimeters (ZDC).  The
number of charged particles measured with the BBCs
and the neutral spectators measured with the ZDCs are correlated 
with the number of participating nucleons, thus together providing a  
measure of the centrality.  The event plane, which is a measure of 
reaction plane, is determined using
the two BBCs at $|\eta|$ = 3.1 $\sim$ 3.9, 
where each counter consists of 64 photo-multiplier tubes 
(PMT's) with quartz Cherenkov radiators in front, surrounding 
the beam pipe. The elliptic axis of the event
plane $\Phi_{\rm measured}$ is calculated by the angle weighted 
with the PMT amplitude using the second
harmonic moment as described in refs.\cite{phx200304,Poskanzer:1998yz}. 
The measured event anisotropy is corrected for a finite
resolution of the measured event plane. The estimated event  
plane resolution $\sigma_{\rm RP}$ = 
$\langle\cos(2(\Phi_{\rm measured}-\Phi_{\rm RP}))\rangle$ 
is 0.3 on average, with a maximum of $\sim$0.4 in the mid-central 
collisions. The corrected $v_2$ is calculated via the formula, 
$v_2$ = $\langle\cos(2(\phi -\Phi_{\rm measured}))\rangle$ / $\sigma_{\rm RP}$.
The phase space used for the determination of the event plane 
for this analysis is 3 $\sim$ 4 units away from the 
mid-rapidity, while the inclusive photon and the 
identified $\pi^0$ are measured at $|\eta|$ $<$ 0.35. 

\begin{figure}
\includegraphics[width=1.0\linewidth]{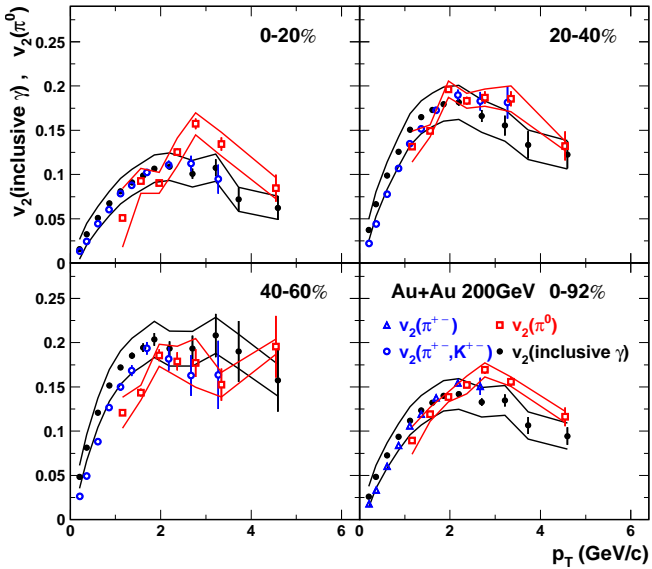}
\caption{\label{fig:fig1} (Color online)
The measured $v_2$ of inclusive photon ($v_2^{inclusive~\gamma}$, 
solid circle) and $v_2$ of $\pi^0$ ($v_2^{\pi^0}$, open 
square) for 4 centrality selections.  The statistical (vertical error 
bars) and systematic errors (lines) are plotted separately. 
The highest $p_T$ point corresponds to 4-6~GeV/${\it c}$.
Charged pion data are from previous measurements \cite{phx200304}.
}
\end{figure}

The photon identification 
and the $\pi^0$ reconstruction are performed in the same way 
as presented elsewhere \cite{phx200302}.
The photon candidate clusters for both inclusive 
photon and $\pi^0$ measurement are first selected 
by their times-of-flight and the corresponding shower 
profiles in the electromagnetic calorimeter (EMCal). 
Neutral pions are reconstructed via $\pi^0$ 
$\rightarrow$ $\gamma\gamma$ decay channel with 
an invariant mass analysis of $\gamma$ pairs.
An additional energy asymmetry cut, 
$|E_{\gamma 1}-E_{\gamma 2}|/(E_{\gamma 1}+E_{\gamma 2})<$
0.8 is applied to the pairs of photon candidates in the 
$\pi^0$ reconstruction. The combinatorial background is 
estimated and subtracted by mixing pairs from different 
events with similar centrality, z-vertex position, and 
event plane orientation. The background is normalized
in a region outside the $\pi^0$ mass peak for each bin in 
relative angle with respect to the measured event plane direction. 
A typical signal over background ratio is
about 1 to 1 at $p_T$ = 3~GeV/${\it c}$ in mid-central collisions
(20-40$\%$ centrality). The $v_2$ of $\pi^0$ is calculated 
from the azimuthal distribution 
after the combinatorial background is subtracted for each 
centrality and $p_T$ bin. For the inclusive photon analysis,
the charged particle contamination in the sample of the 
photon candidate cluster is identified by 
associating the photon candidates with charged particle 
hits in the pad chamber (PC3) directly in front of the EMCal. 
The fraction of photon candidates removed by this charge
veto cut is about 15-25$\%$ depending on centrality. 
The effect of hadron contamination on the measured $v_2$ 
of inclusive photons is estimated by varying the size of
the charged particle association window in the PC3, 
and no significant effect is seen. Neutron and anti-neutron
contamination and off-vertex photons in the identified photon 
sample are studied with full detector Monte-Carlo simulation. 
The correction for these effects is applied to the data;
it is 2$\%$ relative to the measured $v_2$ at 2~GeV/${\it c}$ 
and negligible at 4~GeV/${\it c}$. 
The systematic error includes the effects from the $\pi^0$ 
and photon identification cuts and from the event plane 
determination : 5$\%$ for $\pi^0$ and 5$\%$ for
photon identification and 5-10$\%$ for event plane
determination given by the error on the correction factor 
from the finite event plane resolution. The analysis includes both a
minimum-bias sample (30M events) and a Level2 trigger
sample (equivalent to 55M events), where the Level2 algorithm is 
described in~\cite{phx200500}. 

Figure~\ref{fig:fig1} shows the measured $v_2$ of
$\pi^0$ and inclusive photons as a function of $p_T$ for
different centrality selections. Data are compared with
previous measurements of charged pions \cite{phx200304}.
The $p_T$ and centrality dependences of both the $\pi^0$ and 
the inclusive photon $v_2$ is consistent with that of 
other mesons \cite{phx200304}.  The $v_2$ values are 
significantly above zero up to the highest $p_T$ points. 
The non-zero $v_2$ of $\pi^0$ up to
the highest $p_T$ cannot be explained by flow effects 
alone, but may be attributed to jet quenching and/or 
quark coalescence (recombination).

\begin{figure}
\includegraphics[width=1.0\linewidth]{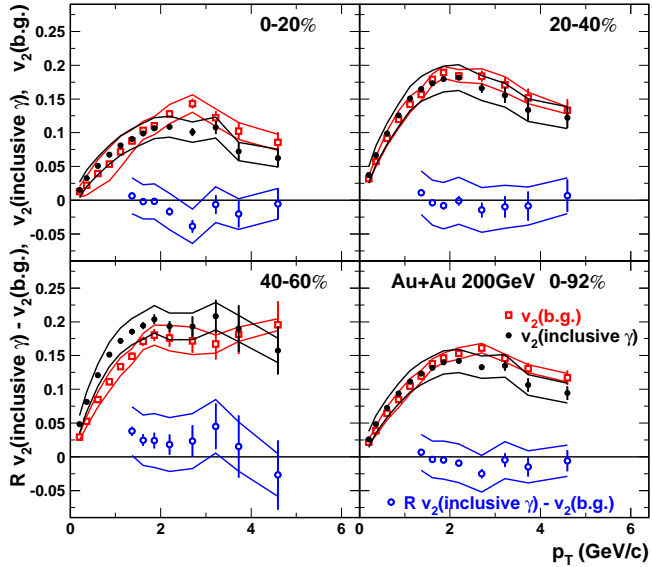}
\caption{\label{fig:fig2} (Color online)
The measured $v_2$ of inclusive photons ($v_2^{inclusive~\gamma}$, 
solid circle) and expected photon $v_2$ from hadronic decay 
($v_2^{b.g.}$, open square). 
A subtracted $v_2$ quantity $R v_2^{inclusive~\gamma} - v_2^{b.g.}$ 
is plotted at the bottom of each panel (open circle), where
$R=(N_{direct~\gamma}+N_{b.g.})/N_{b.g.}$.  The quantity 
corresponds to a product of the direct photon $v_2$ and a 
positive factor $R-1$, ($v_2^{direct~\gamma} (R-1)$). 
}
\end{figure}

Figure~\ref{fig:fig2} compares for different centralities 
the $v_2$ of inclusive photons 
with the expected photon $v_2$ from hadronic decays.
The expected photon $v_2$ from hadronic decays ($v_2^{b.g.}$) 
is calculated by Monte Carlo simulation with the measured 
$v_2$ of $\pi^0$ and other hadronic sources of photon.  
The relative yield of other sources (mainly $\eta$) is 
about 20$\%$ of the total hadronic decay photons, which corresponds 
to about 4$\%$ relative contribution in $v_2$ at 1~GeV/${\it c}$
and negligible at 3~GeV/${\it c}$.  In the simulation, 
we assume that the $v_2$ of 
$\eta$ is similar to the kaon (the closest in mass particle) 
$v_2$ measured in \cite{phx200304,str200401}. 

The $v_2$ of the inclusive photons $v_2^{inclusive~\gamma}$ can 
be expressed as, 
\begin{equation}
v_2^{inclusive~\gamma} = \frac{v_2^{direct~\gamma}N_{direct~\gamma} 
                       + v_2^{b.g.}N_{b.g.}}
                        {N_{direct~\gamma} + N_{b.g.}},
\end{equation}
\noindent
where $v_2^{direct~\gamma}$ is the direct photon $v_2$, 
$N_{direct~\gamma}$ is the direct 
photon yield, and $N_{b.g.}$ is the background photon yield. 
Using the direct photon excess ratio
$R=(N_{direct~\gamma}+N_{b.g.})/N_{b.g.}$, 
previously measured in~\cite{phx200500}, 
one can express the direct photon $v_2$ as:
\begin{equation}
v_2^{direct~\gamma} = \frac{R v_2^{inclusive~\gamma} - v_2^{b.g.}}{R - 1}. 
\label{eq:direct} 
\end{equation}
\noindent 
The bottom data points in each panel of Fig.~\ref{fig:fig2} 
show the difference: $R v_2^{inclusive~\gamma} - v_2^{b.g.}$ 
(the numerator in the above equation), which corresponds 
to a product of the direct photon $v_2$ times a positive 
factor $R-1$, $v_2^{direct~\gamma} (R-1)$.  Alternatively, 
it would be possible to calculate $v_2^{direct~\gamma}$ using 
the measured ratio $R$ ~\cite{phx200500}.  However, we have 
chosen this subtracted quantity in order to show the direct 
photon $v_2$ and its sign, because $R-1$ is measured to be 
small, especially at low $p_T$, and is sometimes negative 
experimentally.
The comparison between $v_2^{inclusive~\gamma}$ and 
$v_2^{b.g.}$ in each panel indicates
that the measured inclusive photon $v_2$ is consistent 
with the expected photon $v_2$ from hadronic decay over
the measured $p_T$ range. The subtracted points are 
close to zero, which is also expected because 
of the lack of the direct photon signal in the measured 
$p_T$ range, where $R$ is close to unity \cite{phx200500}.
The subtraction is especially meaningful where the
measured $R$ value goes above 1.0 at about 4-6~GeV/${\it c}$ and 
higher $p_T$ in central Au+Au collisions \cite{phx200500} ; 
a region where one could extract the direct photon $v_2$.
The measurement indicates that $v_2$ of the direct photon
is small at least in the highest $p_T$ (4-6~GeV/${\it c}$)
range in central Au+Au collisions. 
However, some hidden important trends (slightly negative 
or positive $v_2$ of direct photon) as a function of 
$p_T$ and centrality could be extracted, once the 
errors on those two $v_2$'s and on the measured $R$ 
are small enough. 
This is because the plotted subtracted quantity needs to 
be magnified by $1/(R-1)$ in order to get the direct 
photon $v_2$.
The extracted direct photon $v_2$ at 4-6~GeV/${\it c}$ is 
-1.5$\%$ with $\pm$6.4$\%$ statistical and $\pm$6.4$\%$ 
systematic errors for 0-20$\%$ central events and 
-2.4$\%$ $\pm$6.7$\%$ (sta.) $\pm$9.8$\%$ (sys.) 
for 0-92$\%$ (minimum bias) events.

\begin{figure}
\includegraphics[width=1.0\linewidth]{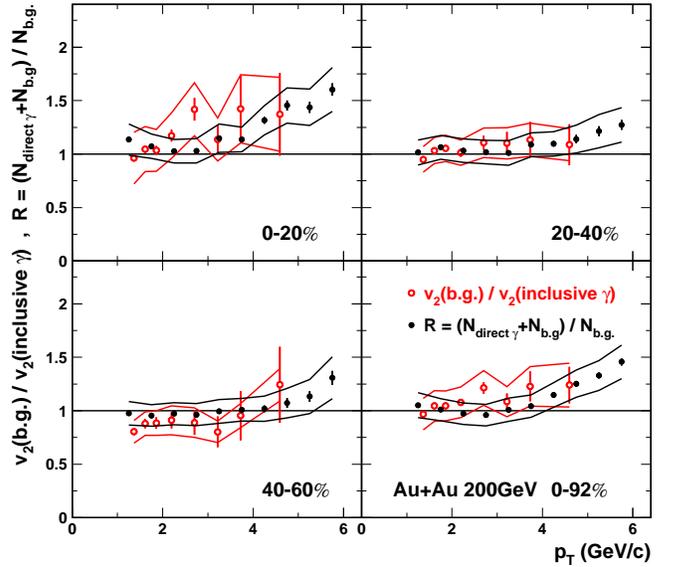}
\caption{\label{fig:fig3} (Color online)
The ratio of the hadronic decay photon $v_2$ over 
inclusive photon $v_2$ ($v_2^{b.g.}/v_2^{inclusive~\gamma}$, 
open circle) compared with the direct photon excess 
ratio $R=(N_{direct~\gamma}+N_{b.g.})/N_{b.g.}$, (solid circle).
}
\end{figure}

Figure~\ref{fig:fig3} shows the 
ratio of $v_2^{b.g.}/v_2^{inclusive~\gamma}$ and a comparison to the 
measured ratio $R$ of the yields from~\cite{phx200500}. 
If the direct photon $v_2$ is assumed to be zero, the ratio 
$R$ should be equal to $v_2^{b.g.}/v_2^{inclusive~\gamma}$  
according to the Eq.~\ref{eq:direct}. 
If the measured direct photon excess comes from the initial hard
scattering, that would correspond to zero $v_2^{direct~\gamma}$, 
then the measured $v_2$ ratio $v_2^{b.g.}/v_2^{inclusive~\gamma}$ gives
a consistent check of the direct photon excess ratio $R$ measurement,
especially where $R$ is significantly above 1.0. The measured $v_2$ 
ratio as a function of $p_T$ and centrality is consistent 
with the conventional relative yield measurement of the direct 
photon excess ratio $R$, but has somewhat larger errors. 

In conclusion, the $v_2$ of identified $\pi^0$ 
and inclusive photons as a 
function of $p_T$ and centrality are measured with 
the PHENIX central arm spectrometer at $|\eta|$ $<$ 0.35 
with respect to the event plane defined at $|\eta|$ = 
3.1 $\sim$ 3.9 in 200~GeV Au+Au collisions at RHIC. 
The $v_2$ of identified $\pi^0$ shows a similar trend
as a function of $p_T$ and centrality compared with other
mesons and has values significantly above zero up to the highest $p_T$ 
point. The measured $v_2$ of the inclusive photons is 
consistent with the $v_2$ of photons from hadronic decays, 
which is furthermore consistent with the absence of 
direct photon signal over the measured $p_T$ range. 
However, the measurement indicates a small direct photon 
$v_2$ for the highest $p_T$ (4-6~GeV/${\it c}$) range 
in central Au+Au collisions. 
The ratio of the estimated photon 
$v_2$ from the hadronic decay over the measured inclusive
photon $v_2$ is also consistent with the direct photon 
excess ratio measured via conventional yields ratio. This
should also imply that the $v_2$ of direct photons is zero 
where the measured direct photon excess ratio $R$ is 
significantly above 1.0. 
The present statistics and systematic accuracy of the 
data from the second year of RHIC running do not allow 
us to explicitly state the magnitude of direct photon $v_2$.
However, the indication of small $v_2$ for direct photons
would favor the naive scenario of direct photon production
from initial hard scattering and its small interaction with 
produced matter in high energy Au+Au collisions. 


%


We thank the staff of the Collider-Accelerator and Physics
Departments at BNL for their vital contributions.  We acknowledge
support from the Department of Energy and NSF (U.S.A.), 
MEXT and JSPS (Japan), CNPq and FAPESP (Brazil), NSFC (China), 
CNRS-IN2P3 and CEA (France), 
BMBF, DAAD, and AvH (Germany), 
OTKA (Hungary), DAE and DST (India), ISF (Israel), 
KRF and CHEP (Korea), RMIST, RAS, and RMAE (Russia), 
VR and KAW (Sweden), U.S. CRDF for the FSU, 
US-Hungarian NSF-OTKA-MTA, and US-Israel BSF.


\end{document}